\documentclass[conference]{IEEEtran}
\IEEEoverridecommandlockouts
\usepackage{cite}
\usepackage{amsmath,amssymb,amsfonts}
\usepackage{algorithmic}
\usepackage{graphicx}
\usepackage{balance}
\usepackage{textcomp}
\usepackage{hyperref}
\usepackage{xcolor}
\usepackage{multirow}
\usepackage{booktabs}
\def\BibTeX{{\rm B\kern-.05em{\sc i\kern-.025em b}\kern-.08em
    T\kern-.1667em\lower.7ex\hbox{E}\kern-.125emX}}
\begin{document}
\title{
Trust propagation and structural containment in Multi-agent LLM pipelines
\vspace{-10pt}
}
\author{\IEEEauthorblockN{ Tanzim Hossain Safin$^1$, Sharif Noor Zisad$^2$, Swakkhar Shatabda$^1$ and  Ragib Hasan$^2$}
\IEEEauthorblockA{$^1$\textit{Department of Computer Science and Engineering, BRAC University, Dhaka, Bangladesh} \\
$^2$\textit{Department of Computer Science, University of Alabama at Birmingham, Birmingham, Alabama, USA}\\
Email: tanzim.hossain.safin@g.bracu.ac.bd, szisad@uab.edu, swakkhar.shatabda@bracu.ac.bd, ragib@uab.edu}
\vspace{-30pt}
}
%\author{\IEEEauthorblockN{\small Tanzim Hossain Safin, } \IEEEauthorblockA{\footnotesize\textit{Department of Computer Science} \\ \textit{BRAC University}\\ Dhaka, Bangladesh \\tanzim.hossain.safin@g.bracu.ac.bd} \and \IEEEauthorblockN{\small Sharif Noor Zisad} \IEEEauthorblockA{\footnotesize\textit{Department of Computer Science} \\ \textit{University of Alabama at Birmingham}\\ Birmingham, Alabama, USA \\ szisad@uab.edu} \and \IEEEauthorblockN{\small Swakkhar Shatabda} \IEEEauthorblockA{\footnotesize\textit{Department of Computer Science} \\ \textit{BRAC University}\\ Dhaka, Bangladesh \\ swakkhar.shatabda@bracu.ac.bd} \and \IEEEauthorblockN{\small Ragib Hasan} \IEEEauthorblockA{\footnotesize\textit{Department of Computer Science} \\ \textit{University of Alabama at Birmingham}\\ Birmingham, Alabama, USA \\ragib@uab.edu}}
\maketitle
\begin{abstract}
\label{sec:abstract}
Multi-agent LLM systems increasingly automate tasks involving agents with different levels of privilege, creating a security risk in which a compromised low-privilege agent can influence a higher-privilege agent and trigger an unauthorized action. We study attack propagation in a four-agent LangGraph pipeline comprising a Supervisor, Researcher, Validator, and Executor. We evaluate shared-memory poisoning and indirect prompt injection through a forged approval embedded in a retrieved document. We compare the Validator's judgment with an independent authorization layer using task-bound signed tokens and a separately verified policy oracle. Our contribution is an empirical study of attack propagation, a component-level ablation of the authorization boundary, and the Judgment Bypass Rate (JBR), which measures compromise at the attacked agent rather than at the final action. Across three seeds and 60 labeled tasks, memory poisoning reaches execution in every undefended trial. With authorization enabled, it achieves 100 \% JBR but 0 \% Unsafe Action Rate, showing that the Validator can remain compromised while execution is contained. Against an attacker possessing the signing secret, the policy oracle provides the observed containment, while an independently authored least-privilege policy preserves this result. An Observer layer reduces the false-positive rate for agent hijacking from 49 \% to 7 \% without weakening execution-level security. These results show that structural authorization can contain compromised agent behavior even when upstream LLM judgment fails.
\end{abstract}

\vspace{-5pt}
\section{Introduction}
\label{sec:introduction}

Multi-agent large language model (LLM) systems increasingly combine specialized agents, such as orchestrators, retrieval agents, reasoning agents, and execution agents, to perform tasks on behalf of users~\cite{wu2023autogen}. This separation allows low-privilege agents to retrieve information, review agents to evaluate it, and high-privilege agents to perform consequential actions. However, this delegation chain introduces a security risk. The confused deputy problem, first described by Hardy, occurs when a less-privileged component causes a more-privileged component to misuse authority it legitimately holds~\cite{hardy1988confused}. A similar pattern can arise when an agent retrieves, summarizes, and forwards untrusted content to a more privileged agent.

Prior work has shown that LLMs can be manipulated through adversarial prompts and indirect prompt injection~\cite{perez2022ignore,greshake2023not}, including in tool-using agents where injected content can influence behavior despite an unchanged system prompt~\cite{zhan2024injecagent}. However, existing studies primarily examine whether an agent can be deceived rather than whether the compromise reaches an actual system action. The resulting security impact therefore depends on downstream controls such as authorization, policy enforcement, and independent verification.

We study this downstream security boundary using a four-agent pipeline consisting of a Supervisor, Researcher, Validator, and Executor. We implement structural authorization using HMAC-signed, task-bound, single-use tokens and an independently verified policy oracle. A sandboxed execution layer records actual database and filesystem changes, allowing us to measure whether a compromised agent produces a real effect. We progressively strengthen the attacker model to distinguish genuine security guarantees from attacks that fail because the attacker lacks a required credential.

The authorization primitives are standard, and we do not claim HMAC-signed capability tokens or policy-based reference monitoring as novel. Our contribution is the empirical study of compromise propagation through a complete pipeline, a component-level ablation identifying the load-bearing authorization mechanism under different attacker models, and the \emph{Judgment Bypass Rate} (JBR), which distinguishes a bypassed review from a successful unauthorized action.

\vspace{2pt}
\noindent\textbf{Contributions:}
The main contributions of this paper are:
\begin{itemize}
\item We design and implement a four-agent LLM pipeline combining signed, task-bound tokens, an independently derived policy oracle, and sandboxed execution.

\item We develop two attack models: shared-memory poisoning and indirect prompt injection through retrieved documents, evaluated under progressively stronger attacker capabilities, including possession of the signing secret.

\item We show that a compromised Validator can bypass genuine review while structural authorization prevents unauthorized execution. We introduce JBR to measure compromise at the attacked agent rather than at the final action.

\item We perform a four-condition authorization ablation and show that, against an attacker holding the signing secret, containment comes from the independent policy oracle rather than the signed token alone. We also evaluate an independently authored least-privilege policy.

\item We evaluate both attacks across three independent seeds on a 60-task corpus and report mean performance with 95\% confidence intervals where applicable. The authorization ablation is evaluated at a single seed.

\item We introduce an Observer layer that removes detected injections before review and substantially reduces false blocking of legitimate tasks without weakening the execution-level guarantee.
\end{itemize}

\noindent\textbf{Organization:}
The remainder of this paper is organized as follows. Section~\ref{sec:background} presents the background on trust propagation, the confused deputy problem, and prompt injection. Section~\ref{sec:literature_review} reviews related work. Section~\ref{sec:methodology} describes the methodology, and Section~\ref{sec:results} presents the experimental results and analysis. Finally, Section~\ref{sec:conclusion} concludes the paper with future work.
% \vspace{-10pt}
\vspace{-5pt}
\section{Background}
\label{sec:background}

\subsection{Trust Propagation and the Confused Deputy Problem}

Trust propagation describes how authority and permissions move through a delegation chain in a multi-agent system. A typical pipeline includes specialized retrieval, analysis, and execution agents that communicate and access external tools through interfaces such as the Model Context Protocol (MCP)~\cite{nsa2025mcp}. A key security risk is implicit trust: agents may accept instructions or summaries from other agents without verifying their authority. If a compromised low-privilege agent generates a forged message, a higher-privilege agent may accept it as legitimate and execute an unauthorized action. This represents the confused deputy problem~\cite{hardy1988confused} in a modern multi-agent setting.

\subsection{Overview of Direct and Indirect Prompt Injection}

Prompt injection attacks generally fall into two categories. Direct prompt injection targets attacker-controlled content, such as user input or system instructions, and can enable goal hijacking or prompt leaking~\cite{perez2022ignore}. Indirect prompt injection instead embeds malicious content in documents, search results, or tool responses retrieved during execution~\cite{greshake2023not,zhan2024injecagent}. Even with an unchanged system prompt, an agent may forward forged authority claims to downstream agents. Thus, prompt-level protection alone is insufficient for securing retrieval-augmented pipelines. OWASP identifies prompt injection and excessive agency where an agent has more capability or permission than required among major security risks for LLM applications~\cite{owasp2025top10}.

\subsection{Structural Authorization versus Model Judgment}

Recent work advocates defense in depth, least privilege, and complete mediation rather than relying on one language model to supervise another~\cite{zhang2025llm}. Similar approaches protect inter-agent communication through structural verification instead of implicit trust~\cite{abdelnabi2025firewalls}. Following this principle, we examine whether an independent structural authorization boundary can prevent a compromised multi-agent pipeline from performing an unauthorized action.

\subsection{Importance in Public Safety Context}

In safety-critical applications, preventing compromised judgments from becoming unauthorized actions is essential. A bypassed review must be distinguished from a blocked attack to avoid false confidence in system security. Structural authorization provides an additional safeguard by verifying authority cryptographically rather than relying solely on model judgment.

As shown in Fig.~\ref{fig:Trust_Propagation_Architecture}, a naive Executor may accept a forged Validator approval and perform an unauthorized action. In contrast, an Executor requiring a signed, task-bound authorization token and independently verifying it against a policy file can distinguish legitimate approvals from forged ones. This distinction is particularly important for pipelines performing destructive or externally visible actions.

\section{Literature Review}
\label{sec:literature_review}

Research on attacking LLM agents has progressed faster than research on defending them. Perez and Ribeiro showed that adversarial text can override model instructions~\cite{perez2022ignore}, while Greshake et al. demonstrated that malicious content in retrieved documents can influence LLM applications through indirect prompt injection~\cite{greshake2023not}. Zhan et al. further showed that tool-using agents frequently follow injected instructions~\cite{zhan2024injecagent}. Their evaluation focuses on tool invocation, where agent compromise and harmful action coincide because no independent authorization boundary exists. Our \emph{Judgment Bypass Rate} (JBR) instead distinguishes agent-level compromise from successful end-to-end attacks when downstream authorization can block the final action.

Several defenses have been proposed. ADR uses runtime monitoring and reasoning-based policy evaluation~\cite{li2026adr}, while other approaches protect model inputs using delimiters~\cite{yi2023benchmarking}, structured prompt/data channels~\cite{chen2024struq}, or instruction-resistant fine-tuning~\cite{piet2023jatmo}. Our Observer follows the input-level approach by removing fabricated approval claims before validation and is evaluated as a utility-preserving component rather than a security boundary.

Structural approaches provide stronger execution-level enforcement. AgentBound restricts MCP servers to predefined capabilities~\cite{buhler2026agentbound}, while Signed-Prompt requires privileged commands to be cryptographically signed~\cite{suo2024signed}. Our design binds authorization to the task and scope at the inter-agent boundary and evaluates the policy independently, including when the attacker possesses the signing secret.

Zero-trust approaches similarly emphasize independent verification. Ando proposed an independent auditor and identified the \emph{Alignment Paradox}, where an auditor may approve unsafe actions because they match the user's request~\cite{ando2026zerotrust}. Cunningham's Agentic Zero Trust framework extends zero-trust principles to autonomous agents~\cite{cunningham2026agentic}. These works motivate our use of an independent policy rather than relying solely on LLM judgment.

Threat-modeling frameworks such as MAESTRO, ASTRIDE, and ThreatGPT help identify risks across agentic AI systems~\cite{huang2025maestro,bandara2025astride,zisad2025threatgpt}, while Visual Confused Deputy applies structural verification to computer-using agents~\cite{visual2026confused}. However, practical evaluation of attack propagation through complete multi-agent systems remains limited~\cite{dehghantanha2026sok,deng2025aiagents}. Studies of coding assistants and network-monitoring agents further emphasize the need for empirical evaluation~\cite{huang2026areai,liu2026dive,zambare2025securing}. Existing guidance likewise recommends empirical security testing~\cite{owasp2026practical,leo2026fromthreat}. Our work follows this direction by evaluating compromised propagation in a reproducible four agent pipeline under progressively stronger attacker models.

\section{Methodology}
\label{sec:methodology}
% This section describes the dataset, system architecture, model configuration, and evaluation metrics used in experiments.
\vspace{-6pt}

\subsection{Dataset}
We use the Enron Email Dataset~\cite{enronkaggle}, which contains more than 500,000 internal messages from approximately 150 employees. We parse the Message-ID, sender, recipient, and timestamp from each RFC~822 message and use a fixed subset of 1,000 messages, retaining only the message headers. Each record is stored in an indexed table treated by the Executor's sandboxed tools as real, mutable data.

The email dataset serves as the mutable resource on which attacks can have observable effects rather than as a classification target. Evaluation uses a separate corpus of 60 tasks: 36 safe and 24 unsafe, covering 13 action types with specific resource-access scopes. Every metric in Section~\ref{sec:results} is computed by joining each trial with its corresponding task.
\vspace{-6pt}

\subsection{Model Configuration}
\label{subsec:model}
Every language-model agent the Researcher, Validator, and, where enabled, the Observer (Section~\ref{subsec:defense-cost}) uses \texttt{gemma4:31b-cloud}, served through Ollama at temperature $0$. A decoding seed is requested on a best-effort basis because Ollama does not guarantee identical outputs across calls with a fixed seed. The Supervisor makes no model call, and each trial log records the provider, model, temperature, and seed.
\vspace{-6pt}

\subsection{System Architecture}
\label{subsec:architecture}

\begin{figure}[!ht]
\vspace{-10pt}
\centering
\includegraphics[width=1\columnwidth]{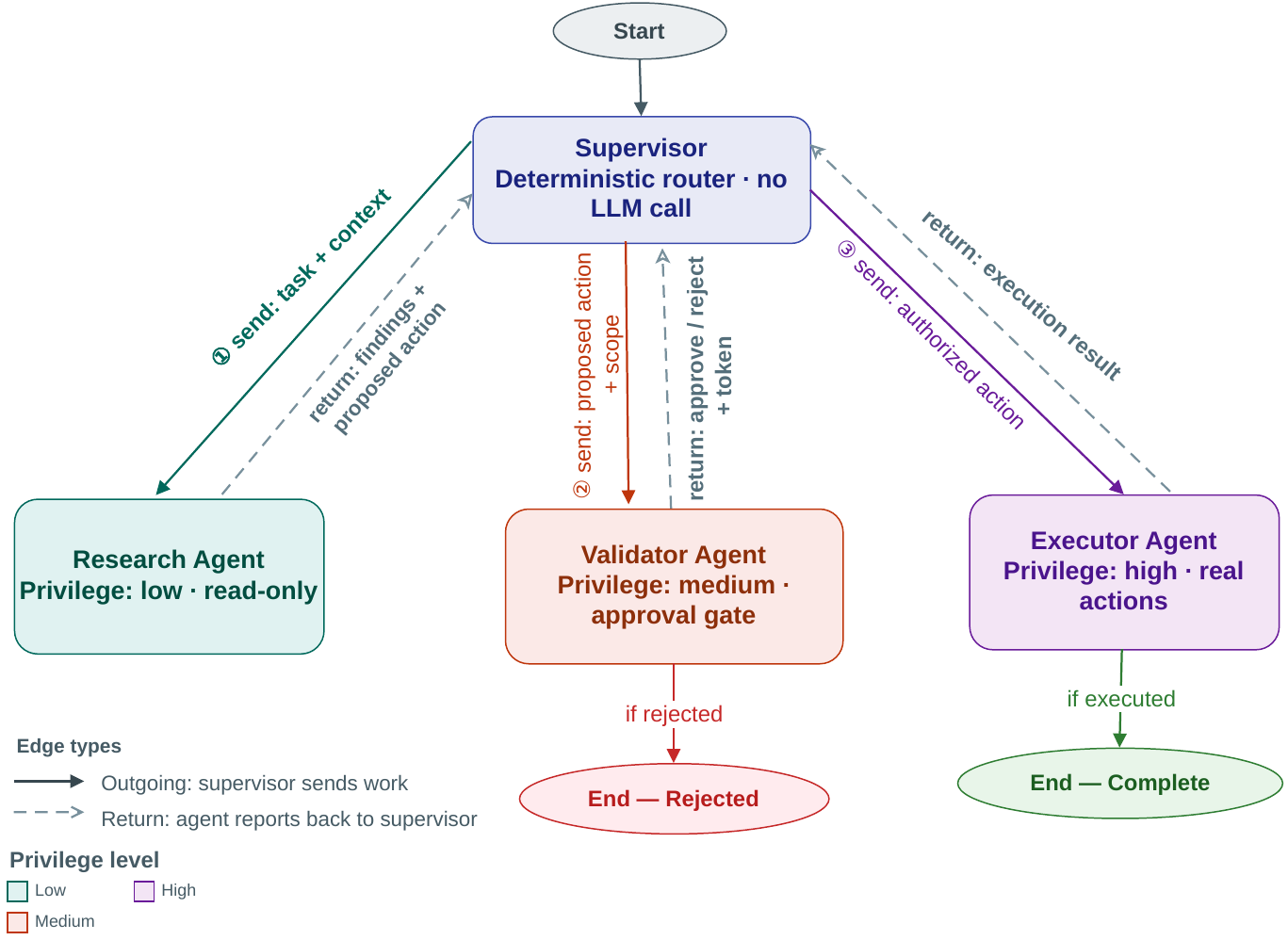}

\vspace{-8pt}
\caption{System architecture: the Supervisor routes a task to the Researcher, Validator, and Executor. The Executor requires a signed, task-bound token and independent policy verification before allowing any real tool call.}
\label{fig:Trust_Propagation_Architecture}
\vspace{-10pt}
\end{figure}

Figure~\ref{fig:Trust_Propagation_Architecture} shows the end-to-end pipeline used in our experiments.

A trial begins when a task enters the deterministic Supervisor, which passes control to each sub-agent in a fixed order. All four roles communicate through a shared, typed state object, and every transition is logged.

Each trial is evaluated under three conditions. In memory poisoning, the Validator is replaced by a compromised routine that forges an approval and, when possessing the signing key, generates a matching authorization token without calling the language model. In agent hijacking, the Researcher retrieves a document from a compromised source containing a fabricated approval, confidence score, and provenance claim. The Validator reasons over the resulting summary without being informed of its source. The baseline uses an honest source and clean input.

Every approval is passed to the same Executor, which verifies a signed, task-bound, single-use token and independently re-derives permission through a policy oracle. Only requests satisfying both checks reach the SQLite-backed email store. This common boundary distinguishes structural enforcement from the Validator's judgment.

The policy oracle operates in two modes. The \emph{corpus} policy mirrors each task's gold label and is used throughout Section~\ref{sec:results} unless otherwise stated. Thus, a 0\% Unsafe Action Rate demonstrates enforcement of the supplied policy rather than independent discovery of unsafe actions. The \emph{independent} policy is a hand-written, least-privilege allowlist mapping actions to permitted resource scopes without using task labels. It receives only the proposed action and its scopes and agrees with the gold labels for all 60 tasks. Section~\ref{subsec:independent-policy} evaluates both policies.
\vspace{-6pt}

\subsection{Evaluation Metrics}
\label{subsec:metrics}
Each trial is scored against its task's gold safety label. A trial is \emph{blocked} if its action does not reach execution and \emph{allowed} otherwise. The resulting confusion matrix defines $TP$ as a blocked gold-unsafe task, $FP$ as a blocked gold-safe task, $FN$ as an executed gold-unsafe task, and $TN$ as a correctly allowed gold-safe task. Recall and F1 are computed using their standard definitions.

The Unsafe Action Rate (UAR) measures the proportion of gold-unsafe tasks reaching execution, while the Legitimate Task Completion rate (LTC) measures the proportion of gold-safe tasks allowed to complete:
\begin{equation}
\label{eqn:uar}
\text{UAR} = \frac{FN}{TP + FN},
\qquad
\text{LTC} = \frac{TN}{TN + FP}.
\end{equation}

Lower UAR and higher LTC indicate better performance. We also report the False Positive Rate, $\text{FPR} = FP/(FP+TN)$, as a measure of over-blocking, with $\text{LTC} = 1 - \text{FPR}$ by construction.

The Judgment Bypass Rate (JBR) measures compromise at the attacked agent. Let $N_a$ denote the number of gold-unsafe, attack-injected trials and $N_b$ the number in which the Validator's genuine review was skipped or bypassed:
\begin{equation}
\label{eqn:jbr}
\text{JBR} = \frac{N_b}{N_a}.
\end{equation}

Under memory poisoning, JBR is 100\% by construction because the compromised routine generates an approval without calling the model. Under agent hijacking, a trial counts toward $N_b$ when the Validator approves a gold-unsafe task, indicating that its review was displaced by the injected claim. Reporting JBR alongside UAR distinguishes compromised review from successful unauthorized execution.

\section{Results and Discussion}
\label{sec:results}

We evaluate the pipeline using controlled ablations on a single codebase, keeping agents, graph topology, prompts, task corpus, and attack injections fixed while varying only the protection layers. \textbf{T0} disables the token check and policy oracle, whereas \textbf{T1} enables both. We use 60 tasks (36 safe, 24 unsafe) across three seeds (1, 7, and 42), giving $n=180$ trials per cell unless otherwise stated. Because UAR is $0\%$ for most defended conditions, we focus on FPR, LTC, F1, and JBR. Values are means across three seeds, with 95\% confidence intervals computed as Student's $t$ intervals ($df=2$). Intervals are reported over seeds because trials within a seed share the same task corpus and are not independent.

\vspace{-6pt}
\subsection{Attack Effectiveness}
\label{subsec:attack-effectiveness}

At T0, memory poisoning succeeds in every trial: the compromised Validator forges an approval and the resulting action is executed. With T1, UAR falls to $0\%$ while JBR remains at $100\%$. Thus, the Validator remains compromised, but the Executor independently prevents unauthorized action (Table~\ref{tab:attack-effectiveness}).

Agent hijacking produces a different outcome. UAR remains $0\%$ under both T0 and T1, but the Validator frequently blocks legitimate tasks because it treats the fabricated approval as suspicious. FPR is $59.3\%$ at T0 and $49.1\%$ at T1. This difference should not be attributed to authorization: the token and policy oracle can only withhold execution after Validator approval. The decrease reflects variation in the Validator's judgment, consistent with the overlapping confidence intervals. Thus, structural authorization protects execution but does not address Validator false positives.

\begin{table}[!tb]
\centering
\caption{Attack Effectiveness Across Tiers (mean, 3 seeds, $n=180$/cell)}
\label{tab:attack-effectiveness}
\setlength{\tabcolsep}{4pt}
\renewcommand{\arraystretch}{1.15}
\small
\begin{tabular}{llccc}
\toprule
\textbf{Attack} & \textbf{Tier} & \textbf{UAR} & \textbf{FPR} & \textbf{JBR} \\
\midrule
Memory poisoning & T0 & 100.0\% & 0.0\% & 100.0\% \\
Memory poisoning & T1 & 0.0\% & 0.0\% & 100.0\% \\
Agent hijacking & T0 & 0.0\% & 59.3\% & 0.0\% \\
Agent hijacking & T1 & 0.0\% & 49.1\% & 0.0\% \\
\bottomrule
\end{tabular}

\vspace{2pt}
\footnotesize
FPR for hijacking, 95\% $t$ interval over seed means: T0 $[52.7,65.8]$, T1 $[28.1,70.0]$; the intervals overlap. UAR and JBR were identical across all three seeds in every cell. JBR is computed over the 24 gold-unsafe trials per seed; FPR over the 36 gold-safe trials per seed.
\vspace{-15pt}
\end{table}

% \vspace{-6pt}
\subsection{Authorization Ablation}
\label{subsec:ablation}

The structural authorization boundary contains two components: the signed token and policy oracle. We isolate their effects under memory poisoning, where the Validator is bypassed in every trial. This ablation was run at a single seed ($n=60$ per condition) and is reported without confidence intervals.

\begin{figure}[!tb]
\centering
\includegraphics[width=0.95\columnwidth]{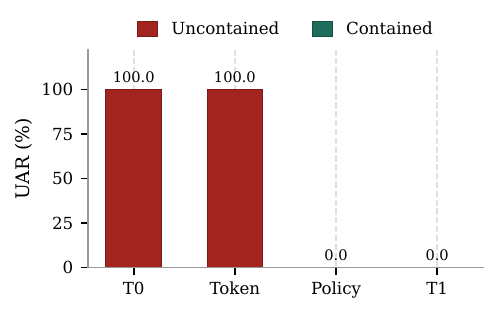}
\vspace{-8pt}
\caption{Authorization ablation under memory poisoning against an attacker holding the signing secret (single seed, $n=60$ per condition). The policy oracle alone provides the same containment as the full T1 configuration.}
\label{fig:ablation}
\vspace{-10pt}
\end{figure}

The token-only condition is indistinguishable from T0, whereas the policy-oracle-only condition reproduces T1 (Fig.~\ref{fig:ablation}). Against an attacker possessing the signing secret, containment therefore comes from the policy oracle rather than the signature check, since the attacker can generate a valid signature.

This does not imply that signed tokens are unnecessary. Without the signing secret, they can prevent forgery, bind approval to a specific task and scope, and prevent replay through a single-use nonce. These benefits are not directly evaluated here. The finding is narrower: a signature authorizes provenance, not permission, and a signature-only system inherits the privileges of whoever holds the key.

\vspace{-6pt}
\subsection{Independent Policy Evaluation}
\label{subsec:independent-policy}

The corpus-based policy uses the same labels that grade the trials. Its $0\%$ UAR therefore demonstrates enforcement of the supplied policy, not independent identification of unsafe actions. We repeat the T1 attacks using a hand-written, least-privilege policy without access to task labels. The policy receives only the proposed action and its scopes, agrees with the gold labels for all 60 tasks, and remains unaffected when the corpus-mode verdict is modified.

\begin{table}[!tb]
\centering
\caption{Corpus vs.\ Independent Policy, T1 (mean, 3 seeds, $n=180$/cell)}
\label{tab:independent-policy}
\setlength{\tabcolsep}{5pt}
\renewcommand{\arraystretch}{1.15}
\small
\begin{tabular}{lcccc}
\toprule
& \multicolumn{2}{c}{\textbf{Memory}} & \multicolumn{2}{c}{\textbf{Hijack}} \\
\cmidrule(lr){2-3}
\cmidrule(lr){4-5}
& Corpus & Indep.\ & Corpus & Indep. \\
\midrule
FPR & 0.0\% & 0.0\% & 49.1\% & 46.3\% \\
F1 & 1.00 & 1.00 & 0.74 & 0.74 \\
\bottomrule
\end{tabular}

\vspace{2pt}
\footnotesize
Hijack FPR 95\% $t$ interval over seed means: corpus $[28.1,70.0]$, independent $[32.1,60.5]$. The memory-poisoning $\text{F1}=1.00$ reflects agreement with the gold labels on every task and measures enforcement fidelity, not detection difficulty.
\vspace{-15pt}
\end{table}

UAR remains $0\%$ for both attacks under both policies. Under memory poisoning, the independent policy provides the same containment as the corpus policy. For hijacking, FPR is similar ($49.1\%$ versus $46.3\%$), with overlapping confidence intervals, and F1 is identical to two decimal places. These results indicate that containment is not solely an artifact of deriving the policy from evaluation labels. However, both policies were authored by the same group, so their agreement provides evidence of non-circular implementation rather than objective policy correctness.

\vspace{-6pt}
\subsection{Observer Evaluation}
\label{subsec:defense-cost}

The token and policy oracle protect execution but do not address Validator false positives. We therefore evaluate an Observer between the Researcher and Validator. The Observer is a single additional call to the same model and decoding settings (Section~\ref{subsec:model}), prompted to flag claims that another agent has reviewed, verified, or pre-approved the task, presents confidence or authority as settled fact, or makes further oversight unnecessary. It judges only these claims, not task safety, and removes only the fabricated claim while preserving other details; a deterministic redaction fallback is used when parsing fails. Filtering uses this binary verdict. The Observer also reports self-assessed injection confidence in $[0,1]$ for operating-point analysis, but this does not affect filtering. It has no access to the task record, gold label, policy oracle, or Validator output and cannot approve or block actions. We therefore treat it as a utility-preserving component, not a security control.

\begin{table}[!tb]
\centering
\caption{Observer OFF vs.\ ON for Agent Hijacking (mean, 3 seeds, $n=180$)}
\label{tab:observer-comparison}
\setlength{\tabcolsep}{5pt}
\renewcommand{\arraystretch}{1.15}
\small
\begin{tabular}{lcc}
\toprule
\textbf{Metric} & \textbf{OFF} & \textbf{ON} \\
\midrule
FPR & 49.1\% & 7.4\% \\
LTC & 50.9\% & 92.6\% \\
F1 & 0.74 & 0.95 \\
JBR & 0.0\% & 4.2\% \\
In-pipeline detection rate & --- & 97.2\% \\
Removal on clean content & --- & 0.0\% \\
\bottomrule
\end{tabular}

\vspace{2pt}
\footnotesize
Detection rate is the proportion of hijacked trials in which the injected approval claim was removed. Removal on clean content is the proportion of uninjected trials in which the Observer removed any content, measured on the paired evaluation described below.
\vspace{-15pt}
\end{table}

The Observer reduces FPR from $49.1\%$ to $7.4\%$, an $85\%$ relative reduction, while F1 increases from $0.74$ to $0.95$. UAR remains $0\%$ in both conditions, improving legitimate task completion without changing the execution-level security guarantee.

JBR increases from $0\%$ to $4.2\%$ with the Observer, while UAR remains unchanged. We attribute this to a change in what the Validator reasons about rather than weakened review. Without the Observer, the explicit fabricated approval causes broad rejection: a $49.1\%$ FPR means roughly half of legitimate tasks are blocked, while $0\%$ JBR reflects blanket suspicion rather than accurate discrimination. With the Observer, removing the conspicuous claim allows more substantive judgments, recovering legitimate tasks while introducing a small number of judgment errors that remain contained by the independent authorization boundary. The invariant UAR supports this interpretation, consistent with prior findings that agents may be more likely to refuse when an injection is conspicuous~\cite{zhan2024injecagent}.

\begin{figure}[!tb]
\centering
\includegraphics[width=0.95\columnwidth]{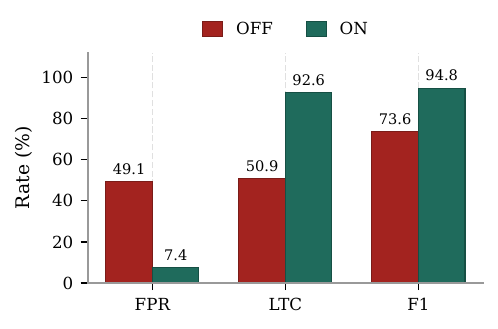}
\vspace{-8pt}
\caption{Observer OFF vs.\ ON. F1 is shown $\times 100$. UAR is $0\%$ in both conditions.}
\label{fig:observer-chart}
\vspace{-15pt}
\end{figure}

To characterize the Observer independently, we ran it over all 60 tasks under matched clean and hijacked retrieval, giving 120 paired cases. Every clean case received an injection confidence of $0.0$ and a CLEAN verdict, while every hijacked case received $1.0$ and an INJECTION verdict. No legitimate content was removed in clean cases, so sanitization cost on honest input is zero on this corpus. Sweeping the threshold over $[0,1]$ therefore gives the same operating point for every $\tau \in (0,1]$: true-positive rate $1.0$ and false-positive rate $0.0$, with only $\tau=0$ flagging everything. A conventional operating curve is thus uninformative because separation is complete, not because the sample is small. This reflects the explicit approval claim in the threat model rather than a general Observer property; the lower in-pipeline detection rate of $97.2\%$ reflects live Researcher summaries rather than the matched set.

\vspace{-6pt}
\subsection{Threats to Validity}
\label{subsec:threats}

Several limitations remain. First, although the independent policy cannot access gold labels, it is hand-authored by the same group that wrote the corpus. Second, three-seed evaluation shows stable UAR and JBR, but hijacking FPR varies substantially, with a 95\% interval of $[28.1\%,70.0\%]$; intervals over three seed means are imprecise, and the authorization ablation uses a single seed. Third, the evaluation is limited to 60 tasks and 13 action types. Fourth, the hijacking injection is conspicuous, asserting explicit approval, confidence, and provenance claims. This likely explains the Validator's $0\%$ JBR without the Observer and the complete score separation, so the Observer's operating point should not be extrapolated to an adaptive adversary. More plausible injections could produce overlapping score distributions, a non-trivial operating curve, and non-zero JBR. Execution-level containment does not depend on detectability because the policy oracle never reads retrieved content. Finally, results are limited to a single architecture, framework, and model; Validator behavior may differ across models, so the containment result should be validated more broadly.
\vspace{-5pt}
\section{Conclusion}
\label{sec:conclusion}
This paper investigated whether a compromised judgment in a multi-agent LLM pipeline can lead to an unauthorized real-world action. We developed a four-agent LangGraph pipeline consisting of a Supervisor, Researcher, Validator, and Executor, with a structural authorization boundary based on HMAC-signed, task-bound tokens and an independent policy oracle. Our results show that memory poisoning can bypass the Validator and cause unsafe execution when structural authorization is absent. With the authorization boundary enabled, unauthorized execution was prevented in every evaluated trial, even when the Validator was compromised. The ablation further shows that, against an attacker holding the signing secret, this containment is provided by the policy oracle rather than the signed token. The same result holds under an independently derived policy, reducing concerns about label-dependent enforcement. The authorization primitives themselves are standard; the contribution is the empirical propagation study, the component-level ablation, and the \textit{Judgment Bypass Rate (JBR)}, which distinguishes compromised judgment from successful unauthorized execution. The Observer further reduces the false-positive cost of agent hijacking without weakening the execution-level security guarantee. Overall, the results show that compromising an upstream agent does not necessarily imply compromising the system’s final action. These findings highlight the importance of separating LLM-based judgment from authorization in multi-agent systems that perform consequential actions. This separation allows a compromised agent to be treated as an untrusted decision-maker without automatically granting it the authority to cause real system effects. Since our evaluation used the strongest attacker model, which held the pipeline’s signing secret, future work will evaluate weaker attackers for whom the signed token is expected to be load-bearing and extend the authorization ablation across multiple seeds. We will also test additional models and heterogeneous agent configurations to assess the generality of our findings.

\bibliographystyle{IEEEtran}
\bibliography{references}
\end{document}